\documentclass[]{aa}  

\usepackage{graphicx}
\usepackage{txfonts}
\usepackage{xcolor}
\usepackage{amsmath}
\usepackage{amsfonts}
\usepackage{natbib}
\usepackage{booktabs}
\usepackage{multirow}
                                
\usepackage{hyperref}
\hypersetup{
    colorlinks=true,
    linkcolor=blue,
    filecolor=magenta,      
    urlcolor=cyan,
    citecolor=teal,
    pdftitle={Identifying correlations among age, expansion, rotation, and shear.},
    }

\begin{document}

   \title{The internal kinematics of local young stellar associations}

   \subtitle{Identifying correlations among age, expansion, rotation, and shear.}

   \author{J. Olivares\inst{1} \and N. Miret-Roig\inst{2,3} \and P.A.B. Galli\inst{4}}

   \institute{
   		Departamento de Inteligencia Artificial, Universidad Nacional de Educación a Distancia (UNED), c/Juan del Rosal 16, E-28040, Madrid, Spain. 
   		\email{jolivares@dia.uned.es}
		\and
		Dep. de Física Quàntica i Astrofísica (FQA), Univ. de Barcelona (UB), Martí i Franquès, 1, 08028 Barc., Spain.
		\and
Institut de Ciències del Cosmos (ICCUB), Univ. de Barcelona (UB), Martí i Franquès, 1, 08028 Barcelona, Spain.
		\and
		Instituto de Astronomia, Geofísica e Ciências Atmosféricas, Universidade de São Paulo, Rua do Matão, 1226, Cidade Universitária, 05508-090 São Paulo-SP, Brazil
}
   \date{}

 
  \abstract
   {The local (<200 pc away) young (<50 Myr old) stellar associations (LYSA) provide fundamental evidence for the study of the star formation process in the local neighbourhood.}
   {We aim at exploring robust statistical correlations in the internal kinematics of LYSAs and of these with age.} 
   {We analyse a public data set containing the linear velocity field parameters and expansion ages of 18 LYSAs. We identify the most robust correlations using frequentist and Bayesian methods.}
   {Among the 45 correlations, we identify only four that passed both frequentist and Bayesian criteria, with these four related to radial motions in the Galactic Z direction. We hypothesise several origins for these four correlations and identify the gravitational potential of the Galactic disk as the most likely driving element. It imprinted the observed motions in the parent molecular clouds, and once the stars were formed, it also damped these motions on a timescale shorter than the LYSAs' ages.}
   {The internal kinematics of local young stellar associations contain fundamental information about the star-formation process that is not fully addressed by star-formation theories, in particular, rotation and shear. Although the Galactic potential appears to be the driving force of these correlations, we urge the theoretical community to provide predictions about the internal motions of expansion, rotation, and shear of stellar associations.}


   \maketitle
\nolinenumbers 

\section{Introduction}
\label{intro}

The young stellar associations in the solar neighbourhood offer an excellent laboratory to study the intrinsic and extrinsic factors affecting the star formation process. Due to their young ages (<50 Myr), they are expected to retain traces of the initial condition in which they were formed, and thus, provide insight into the dynamics of their star formation process. In addition, their proximity (<200 pc) allows us to disentangle their internal kinematics with unprecedented detail, particularly after the arrival of the exquisite astrometry of the \textit{Gaia} mission \citep{2016A&A...595A...1G,2023A&A...674A...1G}.

In the \textit{Gaia} era, the internal kinematics of LYSAs have been targeted by several studies \citep[e.g][]{2024MNRAS.533..705W, 2018MNRAS.475.5659W, 2019ApJ...870...32K,2024A&A...689A..11G, 2023MNRAS.520.6245G, 2021A&A...646A..46G, 2021A&A...654A.122G, 2018ApJ...862..138G,2025A&A...694A..60M}. Most of these studies were limited by either working on the observational space or by transforming to the physical space the measurements of those sources having spectroscopic radial velocity measurements, which are magnitude-limited and time-consuming. Recently, \cite{2025A&A...699A.122O} published a catalogue with the largest and most homogeneous compilation of internal kinematics and expansion ages of LYSAs obtained through a Bayesian forward-modelling method that infers the internal kinematics of stellar systems assuming that they follow a linear velocity field (i.e. a first-order Taylor expansion of the velocity field). This catalogue offers the advantage of being homogeneously derived for a set of 18 LYSAs (<200 pc and <50 Myr) whose literature membership lists were unified, cleaned, and scrutinised for the presence of substructures. Thus, in this work, we profit from this catalogue to explore the correlations among the linear velocity field parameters and expansion ages of these LYSAs.

The rest of this work is organised as follows. In Sect. \ref{data}, we describe the details of the catalogue. In Sect. \ref{methods}, we present the classical statistical methods that we will use to explore and identify the most prominent correlations. Then, in Sect. \ref{results}, we show the results of our statistical exploration, paying particular attention to the age trends that we find in our data set. Afterwards, in Sect. \ref{discussion}, we discuss the possible origins of these correlations and age trends. Finally, in Sect. \ref{conclusions}, we present our conclusions and future work.

\section{Data}
\label{data}

We collected the linear velocity field parameters (see the definition of these parameters in Table \ref{table:parameter}; for more details, see Sect. 2.3.2 of \citealt{2025A&A...693A..12O}) and expansion ages inferred by \citet{2025A&A...699A.122O} and reported in their Table 1 and B.2, respectively. These data correspond to the following young stellar associations:  \object{$\epsilon$-Chamaeleontis} (EPCHA), \object{Musca-Foreground} (MuscaFG), \object{$\eta$-Chamaeleontis} (ETCHA), \object{TW Hydrae} A and B (TWA-A and TWA-B), \object{118 Tau} (118TAU),  \object{32 Orionis} (THOR), \object{$\beta$-Pictoris} (BPIC), \object{Octans} (OCT), \object{HSC2597}, \object{Columba} (COL), \object{Carina} (CAR), \object{Platais 8}, \object{Tucana-Horologium} A and B (THA-A and THA-B), and the recently discovered \object{Háap} (90 pc, $25.9\pm4.6$ Myr), \object{Balam} (63 pc, $19.3\pm4.7$ Myr), \object{$\omega-$Aurigae} (OMAU, 36 pc, $40.1\pm5.9$ Myr), \object{Nal} (183 pc, $16.1\pm5.5$ Myr), and \object{Chem} (168 pc, $22.2\pm2.1$ Myr). We note that Háap association is related to 118TAU and THOR, Balam to BPIC, OMAU to COL, and Nal and Chem to CAR \citep[see][for more details]{2025A&A...699A.122O}. Thus, our data set contains 200 data points: 10 parameters for each of the 20 groups, with these groups consisting of the 18 LYSAs, in which two of them have two substructures.

\begin{table}[ht!]
\caption{Parameters of the linear velocity field model}
\label{table:parameter}
\centering
\resizebox{\columnwidth}{!}
{\begin{tabular}{cl}
\toprule
Symbol & Description  \\
\midrule
$\kappa_i$ & Radial rate along $i$ with $i\in[x,y,z]$ \\ 
$\omega_i$ & Rotation rate along $i$ or in the $\perp_i$ plane, with $i\in[x,y,z]$\\ 
$\gamma_i$   & Shear rate along $i$ or in the $\perp_i$ plane, with $i\in[x,y,z]$\\
\bottomrule
\end{tabular}
}
\tablefoot{
The linear velocity field parameters are reported in units of $\rm{m\,s^{-1}pc^{-1}}$.
}
\end{table}

\begin{figure}[ht!]
\centering
\includegraphics[width=0.9\columnwidth]{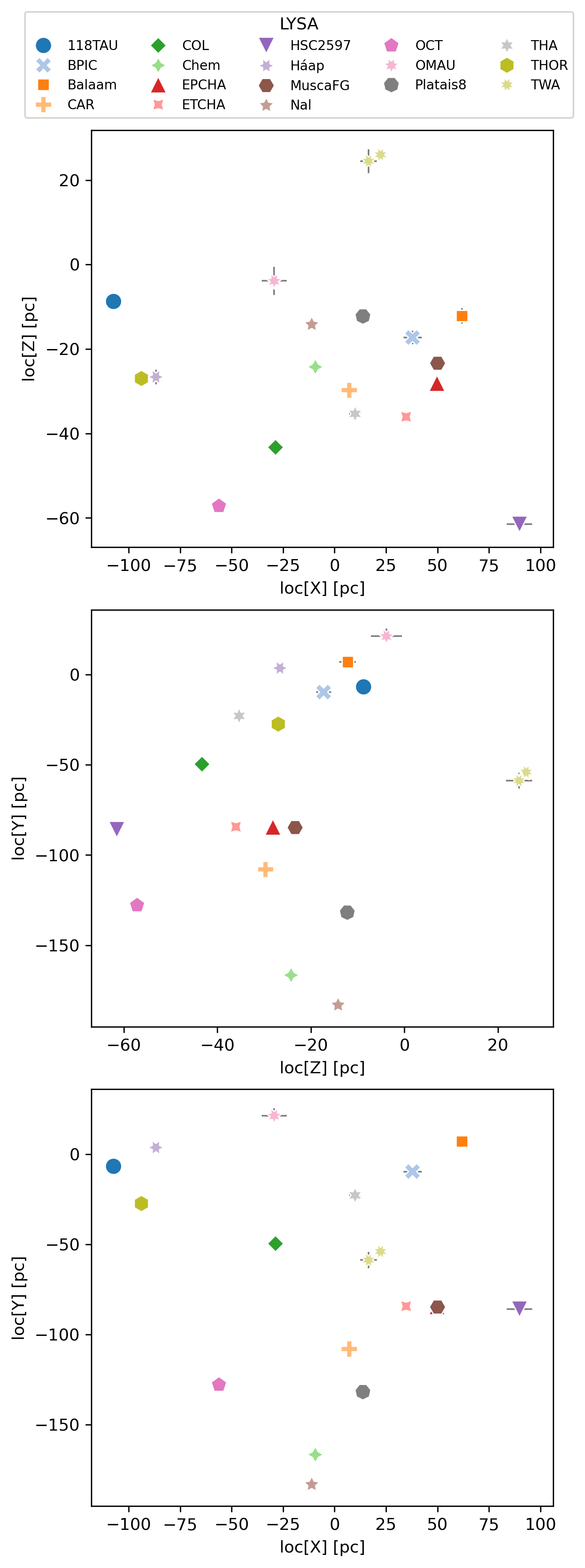}
\caption{Galactic positions of the LYSAs analysed in this work. The symbols and error bars depict the mean and standard deviation of the coordinates' posterior distribution.}
\label{figure:data:xyz}
\end{figure}

Figure \ref{figure:data:xyz} shows the positions, in Galactic coordinates, of the stellar associations whose data we used in this work. As can be observed, the LYSA are contained within a radius of 200 pc around the Sun and are predominantly skewed towards negative values of Y and Z.

Concerning possible biases, we notice the following. First, Table B.2 by \citet{2025A&A...699A.122O} contains the kinematic parameters inferred from several membership lists of the same stellar systems; thus, to avoid biasing our results from repeating non-independent results of the same stellar system, we only use the parameters corresponding to the final membership selection reported by the authors in their Table 1. Second, although the literature membership lists were cleaned of the most probable binary sources \citep[see Sect. 3.2 by][]{2025A&A...699A.122O}, unresolved tight binaries may have influenced the kinematic parameters inferred by the previous authors. However, we notice that the influence of the remaining unresolved binaries can be safely assumed to contribute to the random isotropic increase of the velocity dispersion and thus, their influence on the ordered motions of shear, rotation, and radial expansion or contraction can be assumed to be negligible. Third, here we work with the kinematic parameters inferred by the aforementioned authors based on  their linear velocity field model, which, contrary to the Bayesian expansion age model, assumes neither expansion nor contraction and imposes only a weakly informative normal prior on the $\kappa_i$ parameters \citep[centred at zero with a dispersion of 100 $\rm{m\,s^{-1}pc^{-1}}$; see Table 1 of][]{2025A&A...693A..12O}. Fourth, the spatial distribution of our set of LYSAs (see Fig. \ref{figure:data:xyz}) is restricted to the solar neighbourhood and particularly biased towards the negative sides of the Y and Z galactic directions. Thus, the conclusions that may be drawn from this data set cannot be generalised to the Galactic context nor the extragalactic one. Finally, the age domain of our set of LYSAs goes from 7 Myr to 45 Myr, which implies that our conclusions, if any, will need to be reviewed when data from systems younger than 7 Myr and older than 45 Myr are available. We will further discuss the implications of  these sources of bias in Sect. \ref{discussion:bias}.

\section{Methods}
\label{methods}
In this section, we present the statistical methods that we use to identify significant correlations and model them. We start by screening the most significant correlations based on Pearson's correlation coefficient and a restrictive p-value threshold. Then, out of the most significant correlations, we select the robust ones based on Bayesian and frequentist methods. We note that, given the pioneering origin of this work, our analysis is merely exploratory. 

First, we analyse correlations among the parameters of the linear velocity field and the expansion age of the stellar systems using Pearson's correlation coefficient \citep[implemented in the \texttt{pearsonr} function within SciPy,][]{2020SciPy-NMeth}. We use two-sided null hypothesis tests to discard spurious correlations with p-values larger than $\alpha=0.045$. We assume, as the null hypothesis, that the underlying distributions of samples are uncorrelated and normally distributed. At this first screening, we do not take into account the uncertainties in the linear velocity field parameters or age. We identify robust correlations as those rejecting the null hypothesis with a confidence of 2$\sigma$ (i.e. >95.45\%). In the following, we refer to this method as our frequentist criterion.

Then, we perform a Bayesian hypothesis contrast by modelling the most significant correlations with parametric functions of increasing complexity. We use polynomial models of degree zero (P0) and one (P1) that take into account the uncertainties of the linear velocity field parameters and age (in both coordinates through a bivariate normal likelihood in which the uncertainties enter as the standard deviations in the two-by-two covariance matrix). We also attempted to fit more complex models, like polynomials of degree two and three and an exponential decay for the age correlations, but the dataset lacks the information to constrain the parameters of these more complex models. The use of these nested models, in which the most complex one reduces to the simplest one under certain parameter values (slope zero in this case), allows for a straightforward model comparison. We assume, as the null hypothesis, the zero-degree polynomial and as the alternative hypothesis the linear one. Thus, we reject the null hypothesis of no correlation when the 95.45\% high-density interval (HDI) of the posterior distribution of the slope parameter does not contain the zero value. In the following, we refer to this method as our Bayesian criterion, and we apply it as a veto step to the frequentist one. As prior distributions for the coefficient parameters, we choose a normal distribution with zero mean and a large standard deviation (50 for the intercept and 10 for the slope, in the corresponding units). This weakly informative distribution centred at zero ensures that the zero value is recovered under no constraining data sets. We fit the previous models using the Bayesian Python package \textit{PyMC} \citep{pymc2023} and we analyse the obtained posterior samples with \textit{arviz} \citep{arviz_2019}. 

Also, we did a frequentist fit to the data of the identified significant correlations, but this time with a frequentist method. We performed these fits using the \texttt{curve\_fit} routine from \textit{SciPy} \citep{2020SciPy-NMeth}. In these fits, we also incorporate the uncertainties of the linear velocity field parameters and age, although only in the ordinate variable (the only one allowed by this method). However, given that the 1$\sigma$ reported uncertainties are based on a linear approximation to the function model around the optimum \citep{2005WR004804}, they can be underestimated, and more importantly, the 95.45\% HDI cannot be derived from them. For this reason, we restrict our analysis to the HDI obtained from the Bayesian fit described above.

\section{Results}
\label{results}

\begin{figure}[ht!]
\centering
\includegraphics[width=\columnwidth]{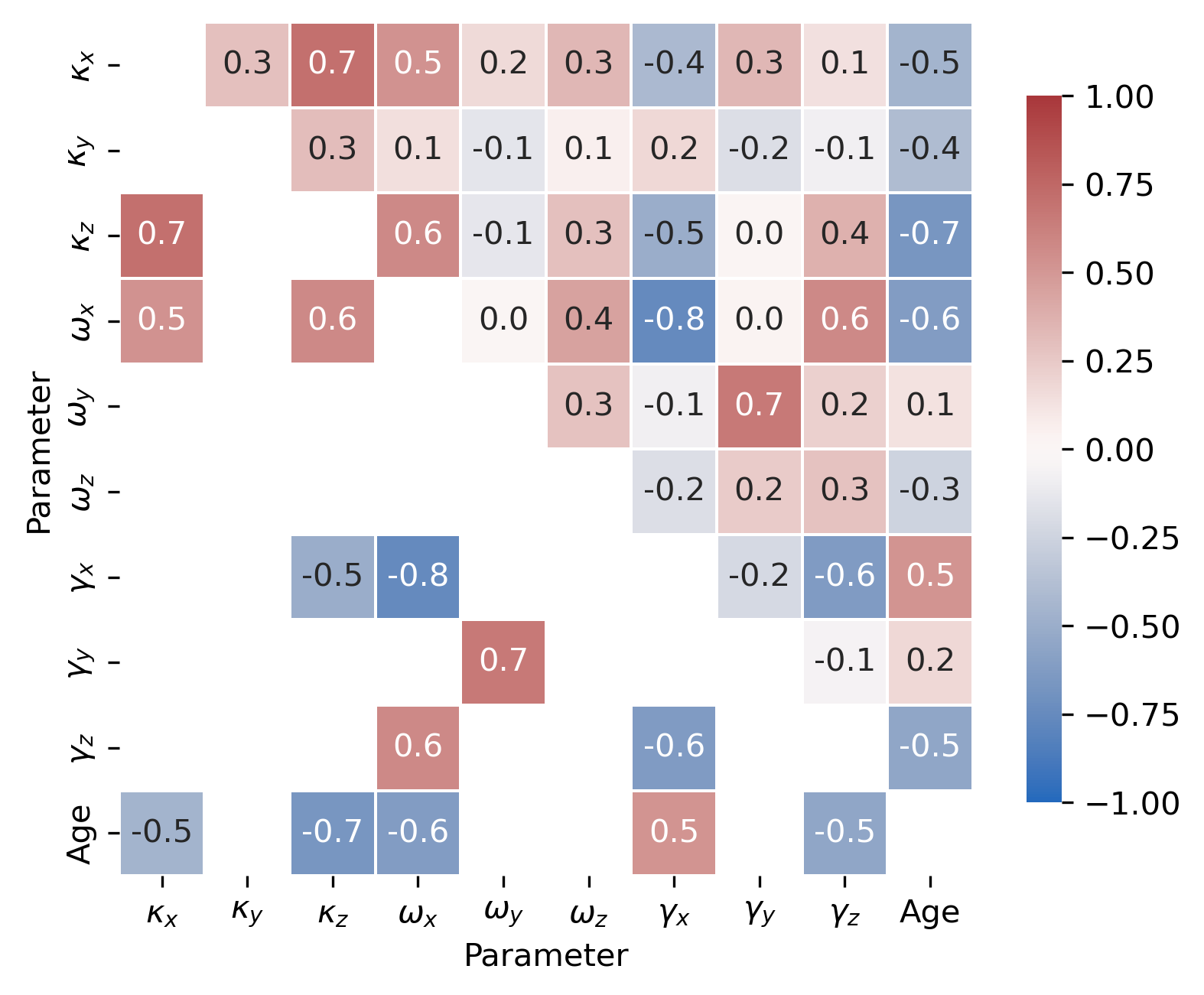}
\caption{Correlation matrix among the expansion age and linear velocity field parameters for our set of 18 LYSAs. The upper triangular part shows all the correlations, while the lower triangular part shows only those passing our 2$\sigma$ criterion. For the sake of simplicity, the trace is not shown.}
\label{figure:correlation_matrix}
\end{figure}

We applied Pearson's correlation coefficient to each pair of parameters in our data set (i.e. the nine parameters of the linear velocity field tensor, $T$, plus the expansion age). Thus, each of these correlation coefficients is computed with a data set of 20 points. Then, we discard those correlations deemed as non-significant according to our 2$\sigma$ frequentist criterion. The results of this analysis are summarised in the correlation matrix shown in Fig. \ref{figure:correlation_matrix}. In this matrix, the upper triangular part shows all correlations, while the lower triangular part only shows those fulfilling our 2$\sigma$ frequentist criterion. For simplicity, the diagonal (i.e. correlations equal to one) is not shown. 

As can be observed from Fig. \ref{figure:correlation_matrix}, out of the 45 correlations, only 13 fulfilled our frequentist criterion. If we were to use a more conservative criterion of 3$\sigma$, only four correlations would have passed it: $\kappa_x$ vs $\kappa_z$, $\omega_x$ vs $\gamma_x$, $\omega_y$ vs $\gamma_y$, and $\kappa_z$ vs age. We consider this $3\sigma$ criterion to be too restrictive under the current relatively small sample size. For this reason, we stick to our original 2$\sigma$ criterion and, in the following sections, we describe the 13 correlations that survived it. Then, we fitted frequentist and Bayesian polynomial models of degree zero and one to the datasets of these 13 correlations (see Table \ref{table:parameters}). Finally, we applied our Bayesian criterion based on the 95.45\% HDI of the Bayesian P1 model (the ninth column of the aforementioned table). Out of the 13 frequentist robust correlations, only the following four passed our Bayesian veto criterion: $\kappa_x$ vs $\kappa_z$, $\kappa_z$ vs $\omega_x$, $\kappa_z$ vs $\gamma_x$, and $\kappa_z$ vs age. The details of these correlations are presented throughout the rest of this section.

\subsection{$\kappa$ correlations}
\label{results:kappa}

\begin{figure}[ht!]
\centering
\includegraphics[width=\columnwidth]{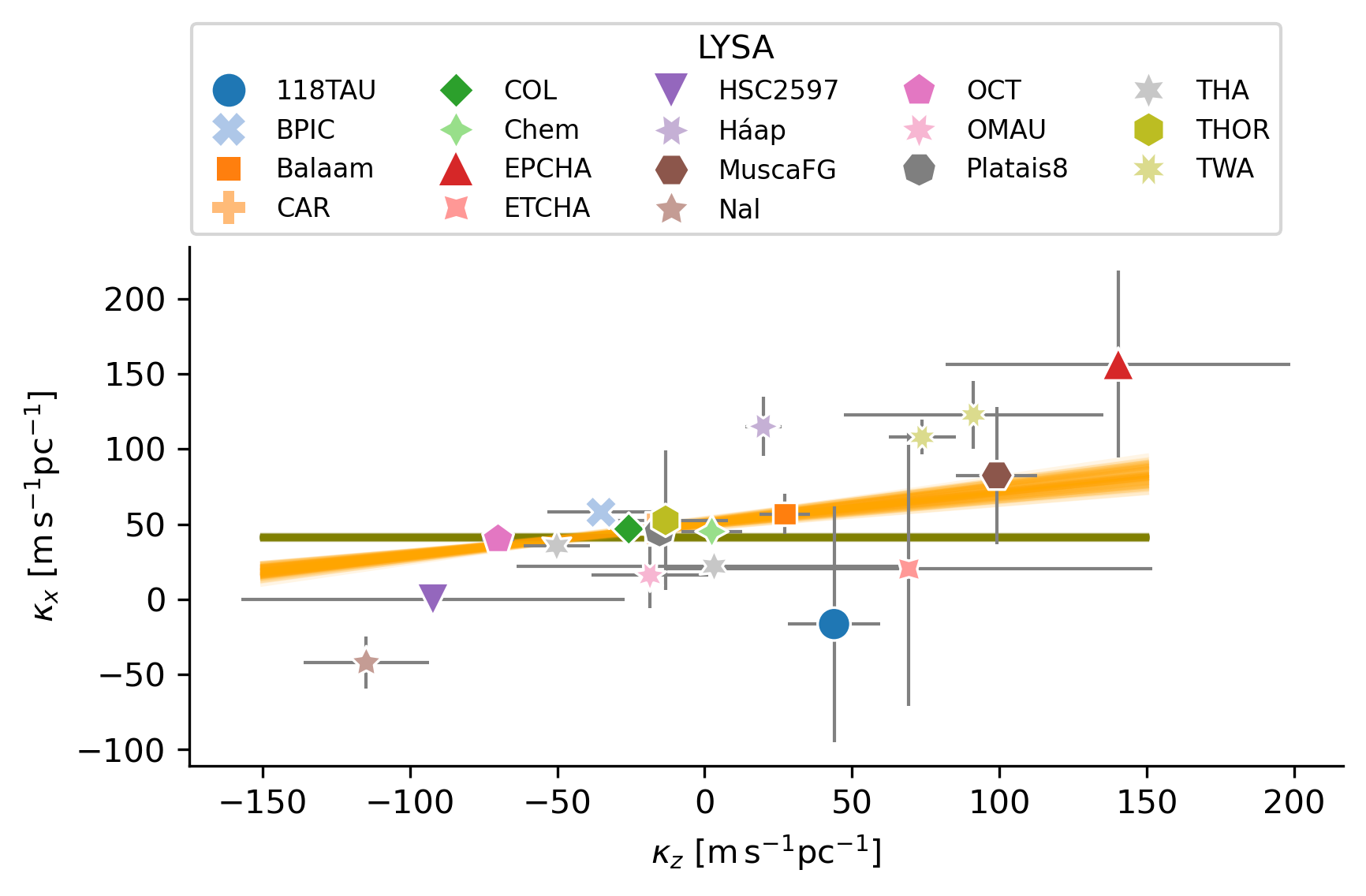}
\caption{Significant correlations for $\kappa_z$. The mean values of the LYSA are shown with coloured symbols and the uncertainties with grey lines. The olive and orange lines show the polynomial fits of degree zero and one, respectively, as obtained from samples of their posterior distributions. We notice that THA and TWA have two substructures, thus their symbols are repeated.}
\label{figure:kappa_z}
\end{figure}

Amongst the $\vec{\kappa}$ parameters, only the correlation between $\kappa_z$ and $\kappa_x$ passed our frequentist criterion. The sample of data points for these correlations is shown in Fig. \ref{figure:kappa_z} together with the Bayesian polynomial models of degree zero (olive lines) and one (orange lines). As can be observed, there is a clear positive linear correlation in the data points that is supported by both criteria (frequentist $\rho=0.7$ and slope HDI [0.155,0.270]). The importance and physical meaning of this correlation will be discussed in Sect. \ref{discussion:kappa}.

\subsection{$\omega$ correlations}
\label{results:omega}

\begin{figure}[ht!]
\centering
\includegraphics[width=\columnwidth]{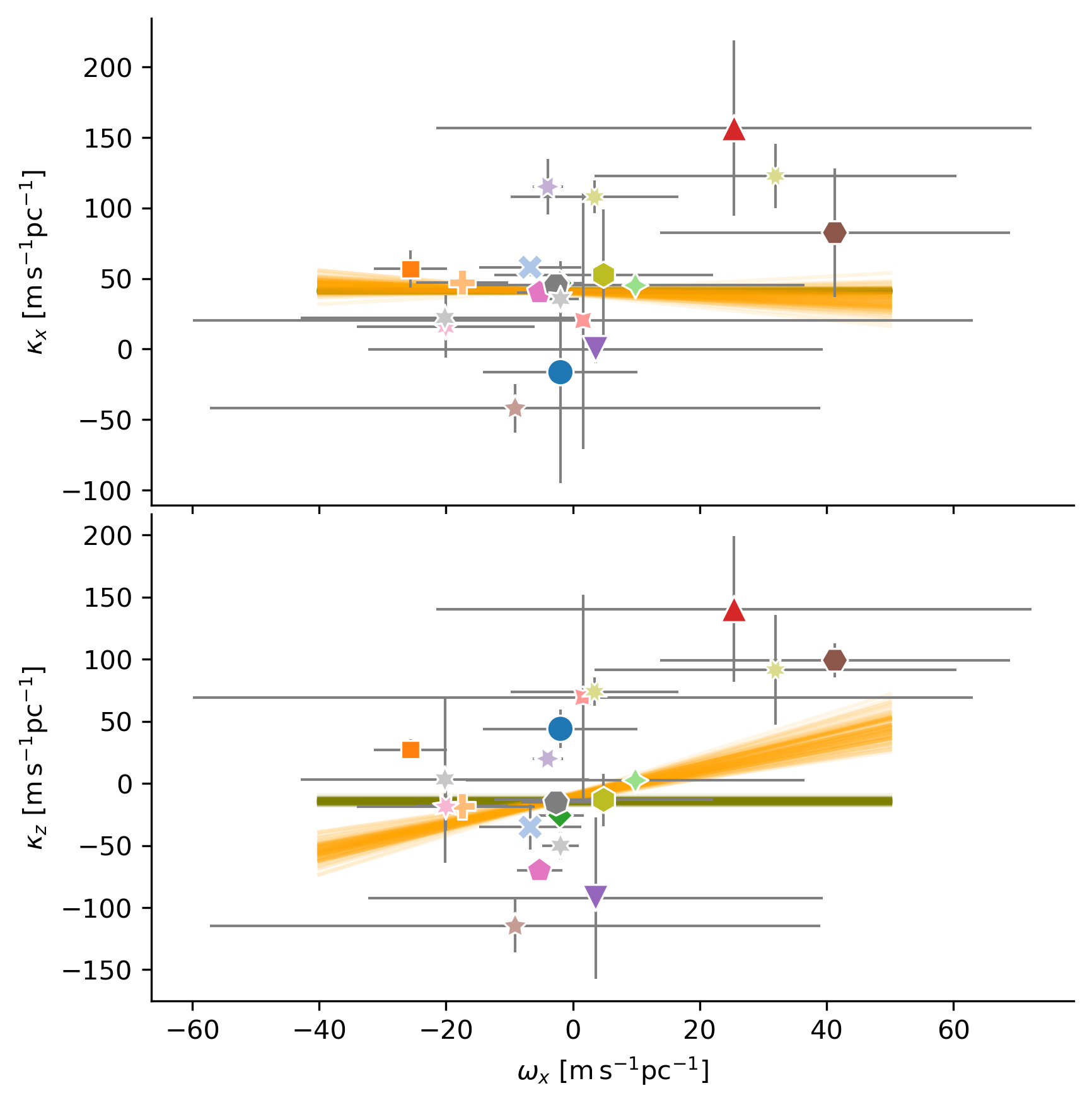}
\caption{Correlations for $\omega_x$. Captions as in Fig. \ref{figure:kappa_z}.}
\label{figure:omega_x}
\end{figure}

The frequentist criterion indicates that the rotation of LYSAs is positively correlated with expansion. In particular, the rotation along the X direction (i.e. in the YZ plane) is positively correlated with expansion along the X and Y directions, with $\rho$=0.5 and $\rho$=0.6, respectively. Figure \ref{figure:omega_x} shows the $\kappa_x$ and $\kappa_z$ data as a function of $\omega_x$ together with the Bayesian polynomials of degree zero (olive lines) and one (orange lines). As can be observed from this figure, only the correlation between $\kappa_z$ and $\omega_x$ survived the Bayesian criterion (slope HDI [0.724,1.514]). The importance and physical meaning of this correlation will be discussed in Sect. \ref{discussion:omega}.

\subsection{$\gamma$ correlations}
\label{results:tau}

The internal shear motion of LYSAs is strongly correlated with the rest of the linear velocity field parameters. We now describe these correlations, direction by direction.

\begin{figure}[ht!]
\centering
\includegraphics[width=\columnwidth]{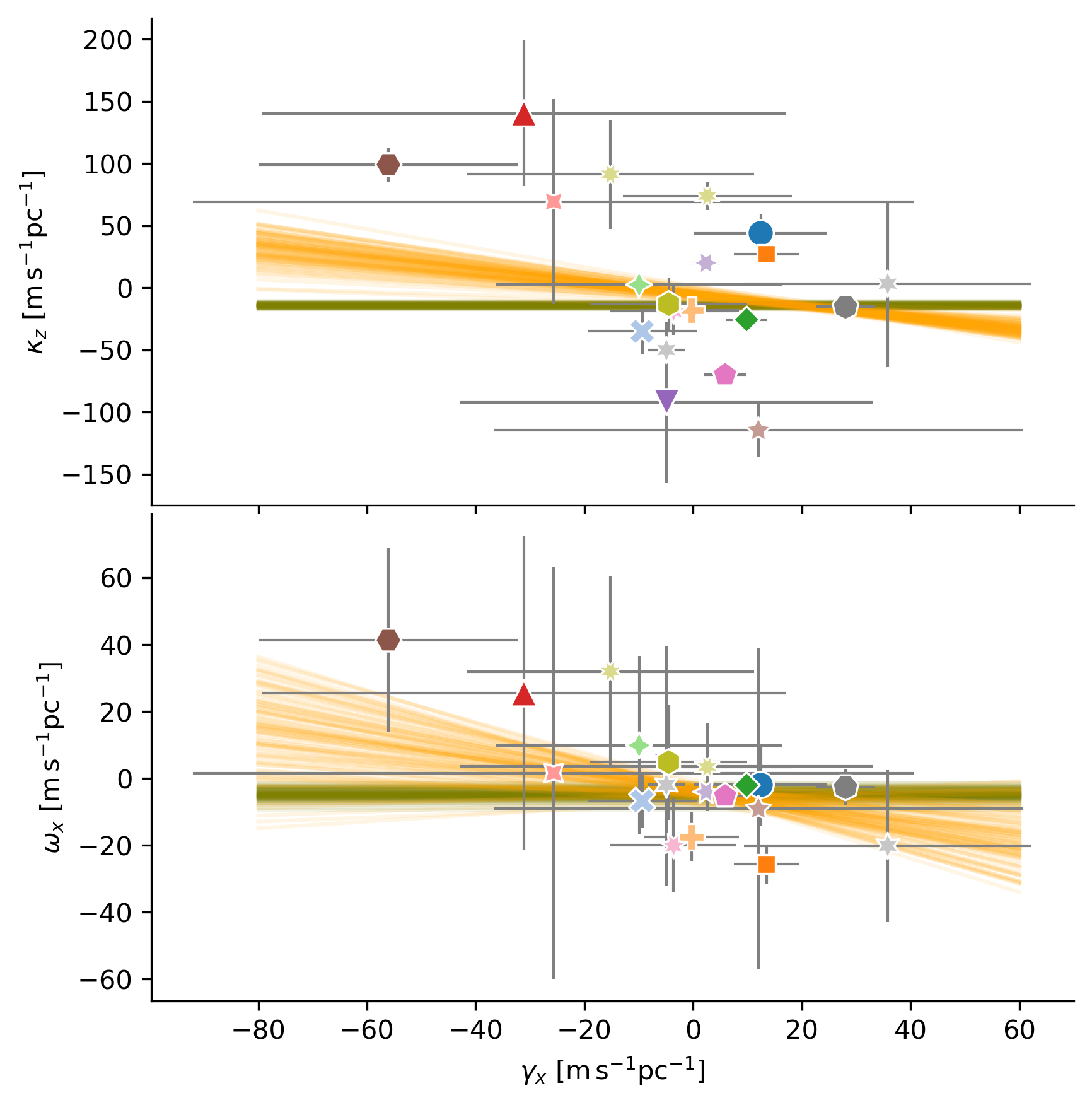}
\caption{Correlations for $\gamma_x$. Captions as in Fig. \ref{figure:kappa_z}.}
\label{figure:gamma_x}
\end{figure}

The shear along the X direction is negatively correlated with expansion and rotation along the Z and X direction, respectively, with frequentist values of $\rho$=-0.5 and $\rho$=-0.8, respectively. The data for these two correlations are shown in Fig. \ref{figure:gamma_x} together with the fitted polynomials of degree zero (olive lines) and one (orange lines). As can be observed from this figure, although the mean values (coloured symbols) are indeed negatively correlated, the large uncertainties on $\gamma_x$ prevent us from drawing stronger conclusions. Indeed, the Bayesian criterion indicates that only in the case of $\kappa_z$ vs $\gamma_x$ the correlation is strong (slope HDI [-0.660,-0.255]). We will further discuss the importance of this correlation in Sect. \ref{discussion:gamma}.

\begin{figure}[ht!]
\centering
\includegraphics[width=\columnwidth]{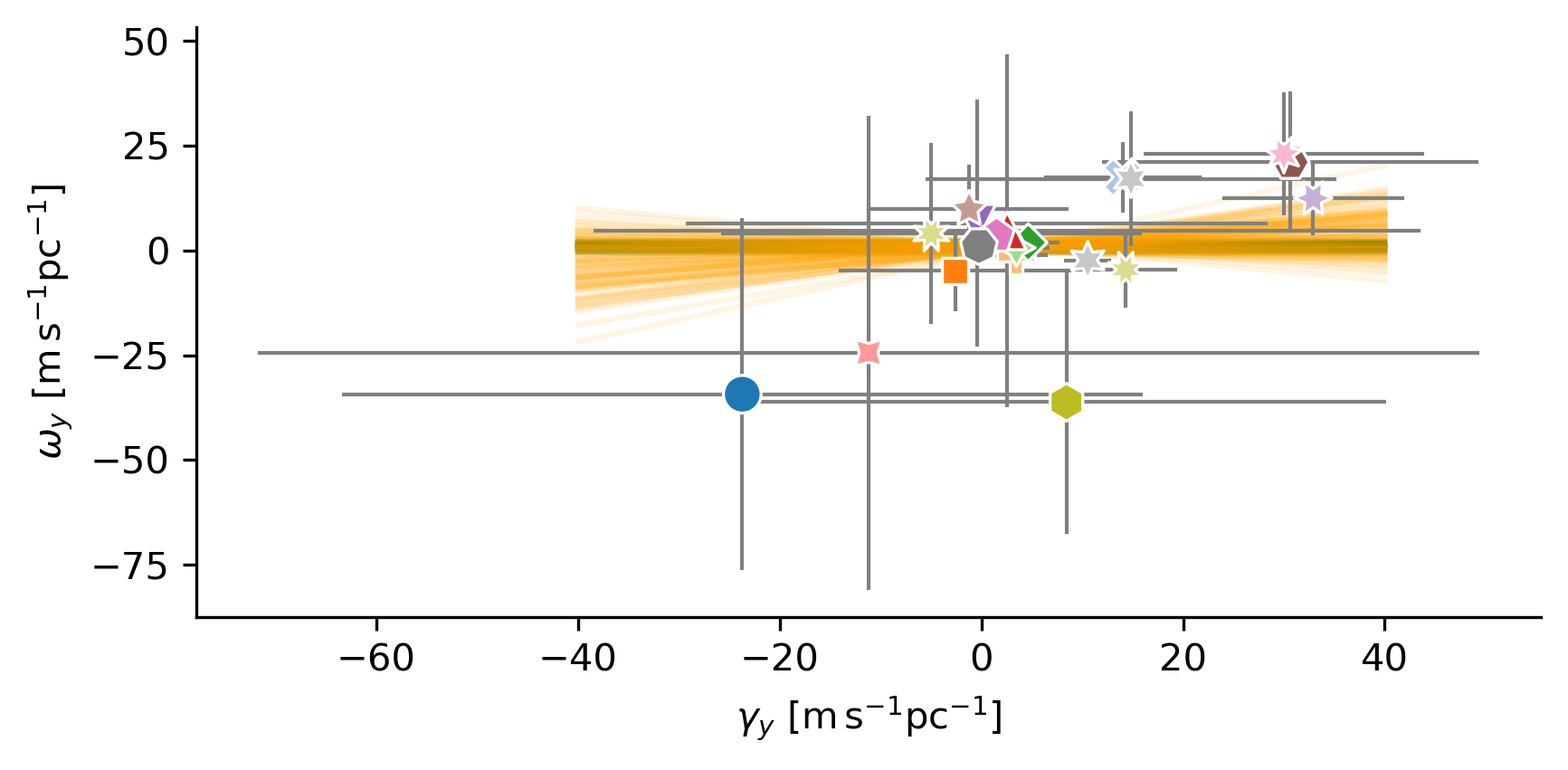}
\caption{Correlations for $\gamma_y$. Captions as in Fig. \ref{figure:kappa_z}.}
\label{figure:gamma_y}
\end{figure}

The frequentist criterion indicates that the shear along the Y direction is only strongly correlated ($\rho$=0.7) with the rotation along the same direction. However, as shown in Fig. \ref{figure:gamma_y}, which contains the data for this correlation as well as the fitted polynomials of degree zero (olive lines) and one (orange lines), this correlation does not survive our Bayesian criterion.

\begin{figure}[ht!]
\centering
\includegraphics[width=\columnwidth]{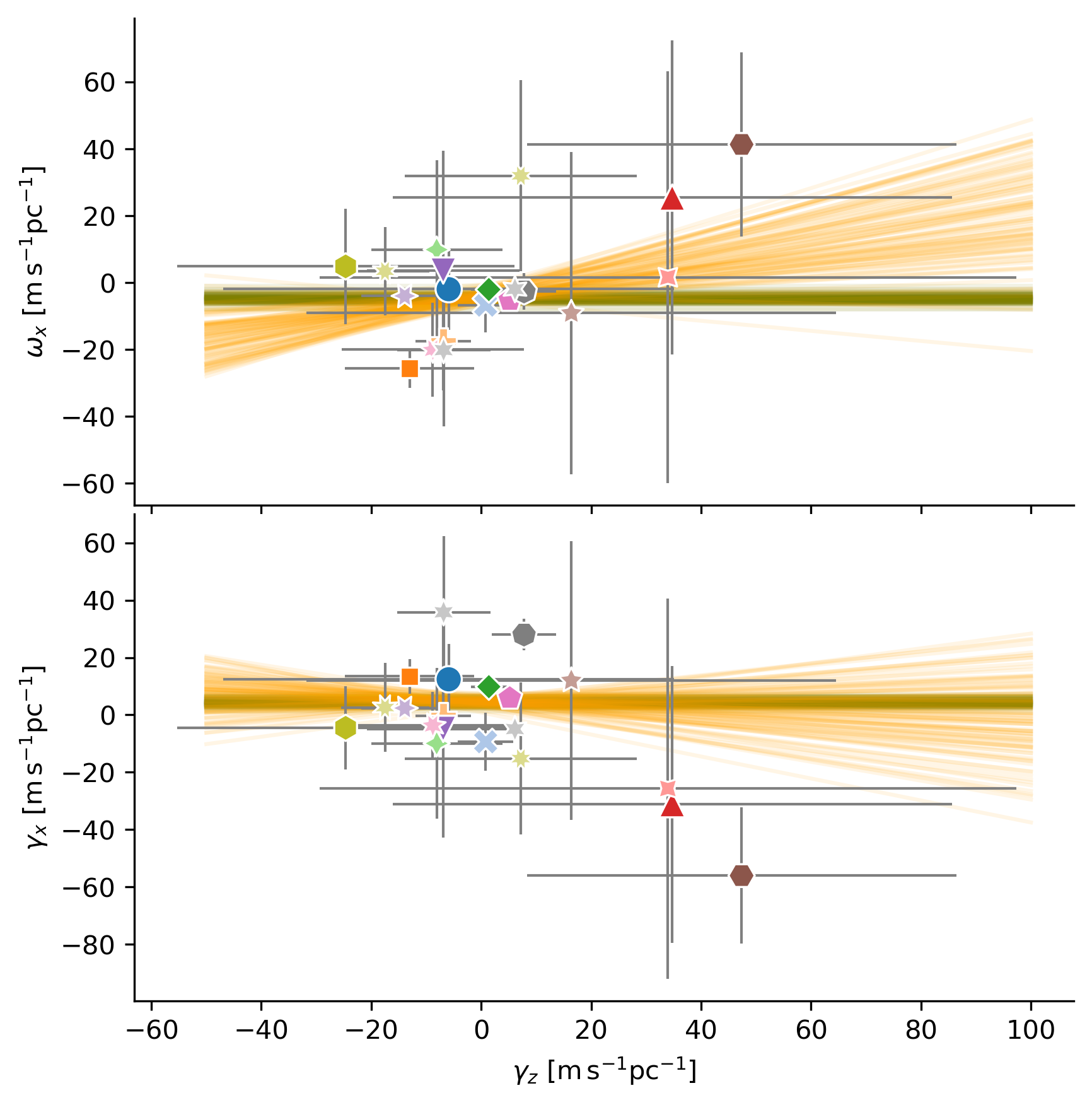}
\caption{Correlations for $\gamma_z$. Captions as in Fig. \ref{figure:kappa_z}.}
\label{figure:tau_z}
\end{figure}

Similar to the previous case, the frequentist criterion specifies that the shear along the Z direction is mildly correlated with the rotation along the X direction ($\rho$=0.6) and mildly anticorrelated with the shear along the X direction ($\rho$=-0.6). However, none of these correlations survived our Bayesian criterion. Figure \ref{figure:tau_z} shows the datapoints for these correlations as well as the fitted polynomial of degree zero (olive lines) and one (orange lines). As can be observed, in both cases, the polynomials of degree one are compatible with those of degree zero. However, it is worth noticing that in the case of $\omega_x$ vs $\gamma_z$, the slope HDI [-0.048,0.514] barely includes zero.

\subsection{Age correlations}

\label{results:age}
\begin{figure}[ht!]
\centering
\includegraphics[width=\columnwidth,height=0.92\textheight]{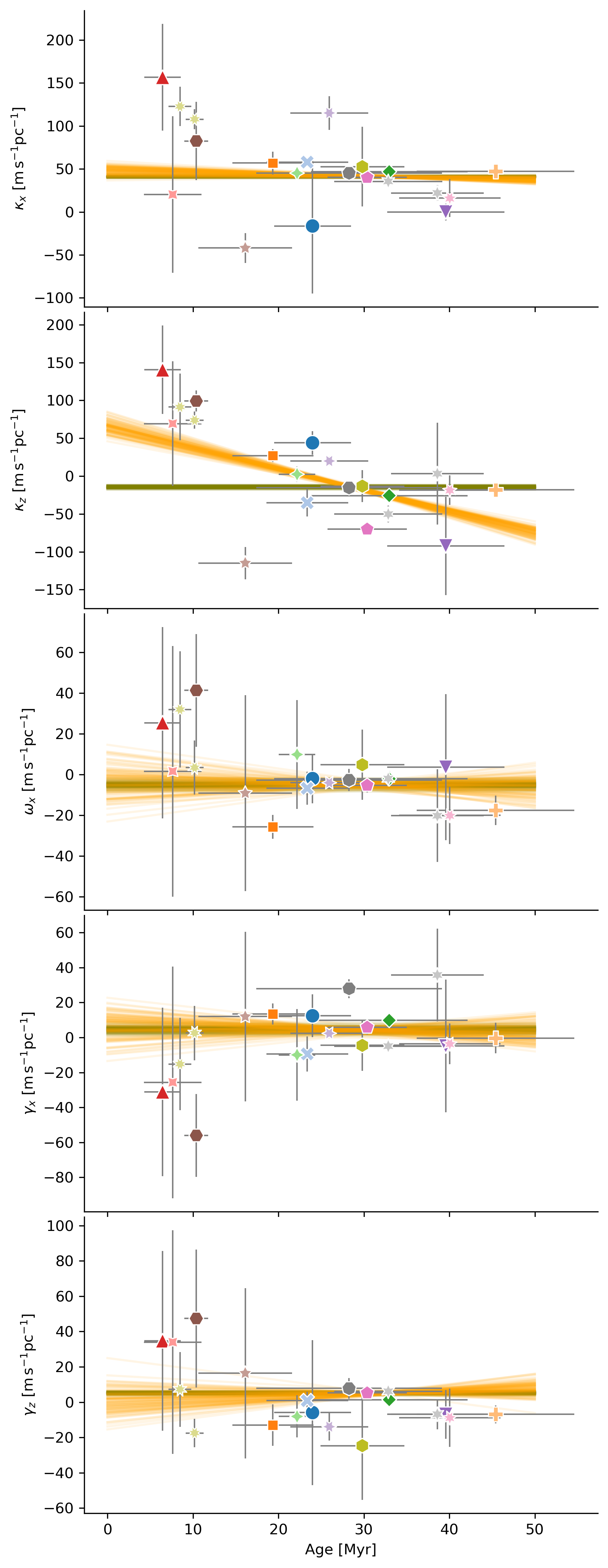}
\caption{Age correlations. Captions as in Fig. \ref{figure:kappa_z}.}
\label{figure:age}
\end{figure}

Figure \ref{figure:age} shows the data for the five age correlations that passed the frequentist criterion together with polynomial fits of degree zero (olive lines) and one (orange lines). As can be observed from this figure, only one of the five correlations survived our Bayesian veto criterion: that of $\kappa_z$ vs age (slope HDI [-3.331,-2.293]). This correlation will be further discussed in Sect. \ref{discussion:age}.

For these five correlations, we attempted to fit an exponential decay, but, in all, except in the case of $\kappa_x$, the result was compatible with zero initial value. In the case of $\kappa_x$ the fitted exponential decay took the following form $\kappa_x(\rm{age}) = 56_{-8}^{+9}\,e^{(-\rm{age}/\tau)}\, \rm[m\,s^{-1}\,pc^{-1}] $ with $\tau=114_{-53}^{+68}$ Myr. This exponential decay is perfectly compatible with the polynomials of degree zero and one; thus, we prefer to continue working with these simpler models.

\section{Discussion}
\label{discussion}

In this section, we discuss the statistical and astrophysical relevance of the robust correlations found in the previous section. First, we discuss the possibility that the correlations result from either a bias in the dataset or in the methods. Then, we proceed assuming there is no bias and we hypothesise an astrophysical origin for each type of correlation (i.e., $\vec{\kappa}$, $\vec{\omega}$, $\vec{\gamma}$, and age). Finally, we provide a global astrophysical explanation for their origin and link this explanation to the star-formation process.

\subsection{Possible biases}
\label{discussion:bias}
\subsubsection{Biased method}
\label{discussion:biased_method}
False positive discoveries may occur when statistical tests are applied to a single dataset, and only those that are significant are reported. This bias is commonly known as p-hacking or data dredging. Several frequentist corrections are available in the literature to mitigate the presence of false discoveries occurring simply by chance. For example, in our methods, the chosen $\alpha$ value of 0.045 may result in 4.5\% of misclassified correlations (i.e. two pairs of correlations).

To estimate the level of false positives, we applied the false discovery control method by \citet{Benjamini-Hochberg_2018}. This method returned only six significant correlations, out of the uncorrected 13 originally found, thus indicating that the false discovery rate of the frequentist method is 53\%. Assuming conservatively that this false discovery rate can be applied to the results of our set of frequentist plus Bayesian tests, then we conclude that out of our final four correlations, two could be spurious by chance. Therefore, it is of paramount importance that future studies, based on independent data sets, confirm or reject the correlations reported in our exploratory analysis.

\subsubsection{Biased data}
\label{discussion:biased_data}
The statistical analysis presented in Sect. \ref{results} showed that the four correlations that survived both our frequentist and Bayesian criteria were all linked to $\kappa_z$. We do not foresee any possible source of bias in the temporal or spatial (see Fig. \ref{figure:data:xyz}) domains of our data that may result in the observed correlations of $\kappa_z$.

However, exploring the data of $\kappa_z$ versus $\kappa_x$, $\omega_x$, $\gamma_x$, and age (see Figs. \ref{figure:kappa_z}, \ref{figure:omega_x}, \ref{figure:gamma_x}, and \ref{figure:age}, respectively), we observe that the points driving these correlations correspond to the LYSAs of EPCHA, ETCHA, MuscaFG, and TWA A and B. With all of them linked to the Scorpius-Centaurus region \citep[Sco-Cen; see for example][]{2025A&A...694A..60M,2023A&A...677A..59R}. 

Thus, one possible explanation for the observed correlations with $\kappa_z$ is that the LYSAs in the Sco-Cen region were formed through a peculiar mechanism of star formation that imprinted internal kinematics signatures different from those observed in the rest of the solar neighbourhood. The validation or rejection of this scenario should await the arrival of more constraining datasets, including the rest of the Sco-Cen groups and groups in other star-forming regions of the solar neighbourhood.

We also explored the possibility that spectroscopic binaries that were not flagged as such by the filtering criteria of \cite{2025A&A...699A.122O} may bias their catalogue, and as a consequence, our results. Thus, for a subsample of the stellar associations (i.e. CAR, Chem, and OMAU), we removed from their list of members those confirmed as binaries by the literature and inferred their linear velocity field parameters following the method outlined by \cite{2025A&A...699A.122O}. With the updated catalogue of linear velocity field parameters, we repeat our correlation analysis, finding the same set of robust correlations identified in Sect. \ref{results}.

The current dataset does not allow us to further test the bias hypothesis. Thus, in the following, we will proceed to discuss each of the robust correlations assuming that there is no bias in our data set.

\subsection{$\vec{\kappa}$ correlations}
\label{discussion:kappa}
The only $\vec{\kappa}$ correlation that survived our both criteria was that of $\kappa_x$ vs $\kappa_z$ (see Fig. \ref{figure:kappa_z}). We notice that this correlation is the strongest that we found, and its strength allowed us to fit polynomials of orders two and three, with both having non-vanishing coefficients that exclude zero from their posterior 95.45\% HDI.

Given the strong and significant $\vec{\kappa}$ correlation between the X and Z directions and the low and non-significant correlation that $\kappa_y$ has with these two directions (see Fig. \ref{figure:correlation_matrix}), we hypothesise that some latent force acting on the Z direction (see Sect. \ref{discussion:force}) perturbs the internal radial motions of LYSA (i.e. in $\kappa_z$), and that this perturbation propagates, through a latent mechanism (more in Sect. \ref{discussion:mechanism}), towards $\kappa_x$. 

\subsection{$\vec{\omega}$ correlations}
\label{discussion:omega}

Concerning rotation, the only correlation that passed both statistical criteria was that between $\kappa_z$ and $\omega_x$. This correlation means that the rotation observed in the YZ plane is positively correlated with the radial motion (contraction or expansion) in the Z direction. Similar to the case between $\kappa_z$ vs $\kappa_x$, we hypothesise that the rotation in the YZ plane results from a latent force acting on the Z direction (see Sect. \ref{discussion:force}) and producing, through some latent mechanism (more in Sect. \ref{discussion:mechanism}).  

\subsection{$\vec{\gamma}$ correlations}
\label{discussion:gamma}

The only $\vec{\gamma}$ correlation that passed both our criteria was that between $\kappa_z$ and $\gamma_x$. Given that $\gamma_x$ indicates a shear motion in the YZ plane, we hypothesise, as in the previous two cases, that a latent force acting preferentially in the Z direction (see Sect. \ref{discussion:force}) results, through some latent mechanism (more in Sect. \ref{discussion:mechanism}), in the shear motion observed in the YZ plane.

\subsection{Age correlations}
\label{discussion:age}

As mentioned in Sect. \ref{results:age}, only the $\kappa_z$ vs age correlation passed our two criteria. Although the rest of the correlations did not pass the Bayesian criteria, we briefly discuss them in the following paragraphs. 

Concerning the temporal evolution of rotation, to the best of our knowledge, this is the first time that such a large compilation of the Galactic 3D components of rotational signals is analysed for nearby young stellar systems. A similar compilation of rotation signals was recently done by \cite{2024A&A...687A..89J}. However, those authors worked with the spin projections in the plane of the sky and in the line of sight as measured from radial velocities and proper motions. We notice that those authors find no correlation between the rotation signal and the clusters' age. Here, although we found a strong frequentist correlation between $\omega_x$ and age, it did not pass our Bayesian criterion.  Nonetheless, the patterns observed in Fig. \ref{figure:age} show hints of a temporal decay. The analysis of this decay should await the arrival of more constraining datasets.

Concerning the temporal evolution of shear motions, to the best of our knowledge, this is the first time that such a compilation is presented and analysed. Although according to our Bayesian criterion, none of these shear motions was correlated with age, the data for $\gamma_x$ and $\gamma_y$ (see Fig. \ref{figure:age}) show hints of a temporal decay. The analysis of this decay should await the arrival of more constraining datasets.  

Concerning the temporal evolution of the radial motions (i.e. expansion or contraction), we observe that although both $\kappa_x$ and $\kappa_z$ appear to have a similar pattern, only that of $\kappa_z$ survived both our statistical criteria. Despite the statistical significance of the correlation, we notice that the LYSAs older than 25-30 Myr have considerably lower expansion than the youngest ones. A strikingly similar temporal trend was reported by \citet{2024A&A...683A..10D} for a considerably larger compilation of stellar systems. However, those authors identify an age decay trend in the ratio of the mean radial velocity to the radial velocity dispersion. According to them, their trend was in agreement with the expectation of violent relaxation and evolution towards virial equilibrium. Although the trend found by those authors is similar to the one observed in our dataset (see, for example, the two upper panels of Fig. \ref{figure:age}), we notice the following differences. First, we analyse the system's expansion rate vector in Galactic coordinates rather than a radial velocity ratio tied to the observed line of sight. Second, our statistical analysis identifies as robust only the correlation between the system's age and the Z direction of the expansion. Therefore, although the explanation proposed by the aforementioned authors may hold, an additional external force may still be producing the observed asymmetry in the $\vec{\kappa}$ correlations with age, preferentially strengthening the one in the Z direction. Thus, we hypothesise that the trend observed in Fig. \ref{figure:age}, has its origin in a latent force acting on the Z direction (see Sect. \ref{discussion:force}) that damps the expansion of LYSAs in that direction.

\subsection{The latent force}
\label{discussion:force}

As discussed in the previous section, the Z direction plays a fundamental role in the identified robust correlations. Thus, there should be a latent dynamical variable (i.e. a force) acting in this direction and resulting in the observed correlations. Due to the spatial constraints of our data, this force may have either a local origin or a global one. 

Concerning the global origin, we consider that assuming the Galactic potential as the external driving force naturally explains the observed correlations of $\kappa_z$ (more in Sect. \ref{discussion:mechanism}). The Galactic potential can be safely assumed as the driving force that acts on all LYSAs alike. Moreover, its strongest component, the Galactic disk, acts preferentially on the required Z direction.

As possible local origins for the driving force, we consider a supernova explosion and a local gravitational perturbation. On the one hand, a supernova explosion at high Z may result in a shock wave with an almost parallel front to the Galactic plane that may have compressed the molecular clouds in the Z direction and thus resulted, through some hidden mechanism (more in Sect. \ref{discussion:mechanism}), in the observed kinematic patterns and correlations. On the other hand, the possibility of a local gravitational perturbation is currently discarded from the results by \citet{2021A&A...646A..67W}. These authors infer the gravitational potential of the Galactic disk in the solar neighbourhood using a Bayesian hierarchical model and \textit{Gaia} DR2 data. Their results indicate (see their Fig. 7 for Z within 100 pc) that, at the spatial distribution of our data ($X\in[-100,100]$ pc, $Y\in[-200,50]$ pc, and $Z\in[-60,30]$ pc, see Fig. \ref{figure:data:xyz}), the inferred gravitational potential is homogeneous and similar to that of the rest of the solar neighbourhood (within an XY radius of 200 pc and Z within 100 pc). We notice that there is a large perturbation in the gravitational potential inferred by the previous authors (see region A4 in their Fig. 8). However, this perturbation occurs at high Galactic altitude (Z = 400 pc). 

The hypothesis of a local origin for the force acting in the Z direction lacks constraining evidence. Thus, in the following, we will assume that this force has a global origin and that it corresponds to the force exerted by the Galactic disk. 

\subsection{The latent mechanism}
\label{discussion:mechanism}

Under the assumptions that our data is not biased, that the internal kinematics of LYSAs associated with Sco-Cen are not different from those of other LYSAs in the solar neighbourhood, and that the latent force acting in the Z direction corresponds to that of the gravitational potential of the Galactic disk, we are now faced with the question about the physical latent mechanism that transforms the previous force into the observed statistical correlations of $\kappa_z$ with age, expansion, rotation, and shear. 

This latent mechanism should relate the force in the Z direction (i.e. the gravitational potential of the Galactic disk) with expansion in the X direction, rotation in the YZ plane, and shear in the YZ plane. In addition, this same mechanism or another one must explain the temporal decay of expansion, rotation, and shear in a timescale of 20-30 Myr.

We consider three not necessarily independent mechanisms for the origin of the $\kappa_z$ correlations. In the first one, which we call the hereditary mechanism, the correlations were already present in the parent molecular cloud and were simply inherited by the newborn stars. In the second one, which we call the nourishing mechanism, the kinematic correlations were created at the birth time of the associations as a by-product of the star-formation process \citep[see, for example,][and references there in]{2017MNRAS.467.3255M,2020ApJ...904...71L}. Finally, in the third mechanism, the correlations resulted from dynamical interactions of the newly formed stars among themselves and with the Galactic potential. However, we consider this third mechanism unfeasible due to the young ages, unbound state, and collisionless nature of stellar associations. Moreover, this third mechanism is ruled out by the results shown in Sect. \ref{results:age}, in which expansion, rotation, and shear decrease rather than increase with age. We now explore the feasibility of the remaining two mechanisms.

In the hereditary mechanism, the molecular clouds, which can be considered classical collisional fluids, acquired the rotation and shear motions due to the Galactic potential, while the radial motions of contraction and expansion (in all directions) may result from their own gravitational collapse. Their viscosity allows them to react, under a compressing force in the Z direction, with shear motions in the XZ and YZ planes, and in turn, these shear motions will result in local vorticity, and thus rotation. This hereditary mechanism is favoured by the observational evidence of the Milky Way and other galaxies. In the Milky Way, the observed vertical velocity gradient, $\rm{-15\pm4\,m\,s^{-1}\,pc^{-1}}$ \citep{2011A&A...525A.134M}, is expected to produce shear motions in the XZ and YZ planes, and thus, also vorticity. We notice that the value of this vertical velocity gradient is similar to the observed shear motions of the youngest LYSAs (see $\gamma_x$ vs Age in Fig. \ref{figure:age}). In other galaxies, the rotational signals of molecular clouds (measured in the plane of the sky) have values also similar to those observed in our LYSAs (see $\omega_x$ vs Age in Fig. \ref{figure:age}): 6-8 $\rm{m\,s^{-1}\, pc^{-1}}$ in M33 \citep{2018A&A...612A..51B} and 18-22 $\rm{m\,s^{-1}\, pc^{-1}}$ in M51 \citep{2020A&A...633A..17B}.  

In the nourishing mechanism, the parent molecular cloud fragments into smaller clumps that afterwards assemble hierarchically with a mass-dependent relation \citep{2017MNRAS.467.3255M,2020ApJ...904...71L}. This hierarchical assembling produces rotation as a consequence of torques between clumps \citep[e.g.][]{2017MNRAS.467.3255M} or as a direct consequence of the star-formation at the merger of two dwarf galaxies \citep{2020ApJ...904...71L}. This mechanism faces several difficulties in explaining the observed correlations with $\kappa_z$. First, the numerical simulations by \citet{2020ApJ...904...71L} and \citet{2017MNRAS.467.3255M}  indicate that rotation is observed in massive and compact stellar systems while it is elusive in the least massive and sparse systems like stellar associations \citep[see, fore example, the discussion of ][about their run D]{2017MNRAS.467.3255M}. Second, although shear motions are not directly mentioned in any of the previous works, they are also expected to appear from the torques between gaseous clumps. Third, the random nature of the turbulent flows that give birth to the clumps veils the explanation for the correlations of rotation and shear with expansion. For these correlations to appear, the torques between clumps must preferentially occur in the YZ plane. Although this asymmetry may arise from the presence of the Galactic disk potential, the previous simulations do not include it. 

Most likely, the previous two mechanisms operate simultaneously. Nonetheless, once the stellar system was formed and imprinted with the observed kinematic patterns of expansion, rotation, and shear, these vanish in a time scale of 20-30 Myr. Concerning expansion, we expect that this motion will face differential forces due to the Galactic potential, with this latter preferentially damping the motion in the Z direction. Nonetheless, we expect that other phenomena, like violent relaxation and evolution towards virial equilibrium, also play important roles \citep[e.g.,][]{2024A&A...683A..10D}. Concerning rotation, the numerical simulations by \citet{2025A&A...699A.196B} predict that the initial bulk rotation signal will disappear on the timescale of two-body relaxation, which for stellar associations is larger than their age. However, our data shows (see Fig. \ref{results:age}) that the initial rotation vanishes after only 20-30 Myr. Thus, other mechanisms not included in the previous authors' simulations must be at play. In particular, we notice that their simulations used the tidal field created by a point mass galaxy, whereas in the case of the solar neighbourhood, the most important contribution to the gravitational field comes from the Galactic disk. Thus, we hypothesise that the latter is responsible for the damping of stellar motions in the Z direction, particularly those of rotation and shear in the YZ plane and expansion in the Z direction.

\section{Conclusions}
\label{conclusions}

\begin{figure}[ht!]
\centering
\includegraphics[width=\columnwidth]{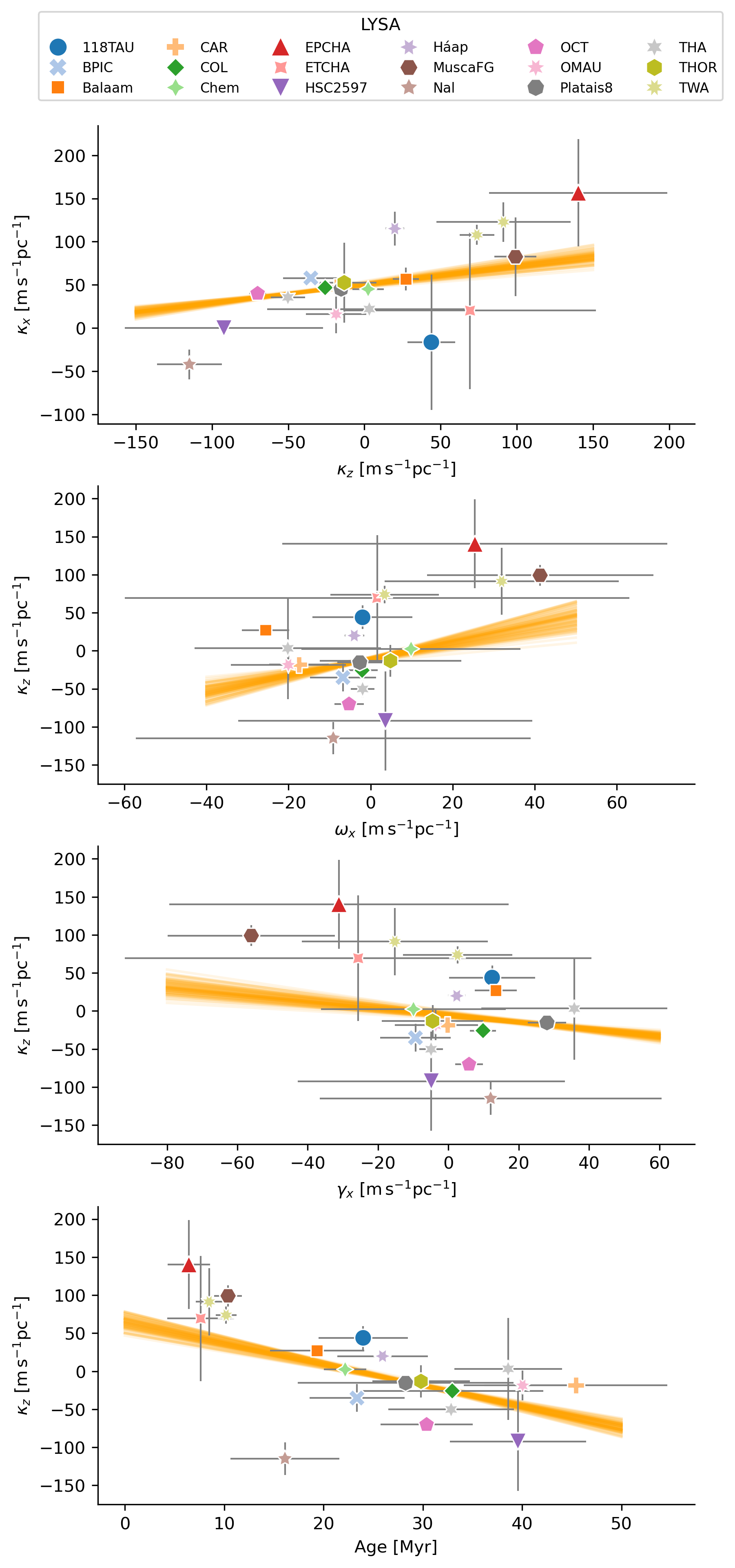}
\caption{Summary of the identified robust correlations. The symbols and grey bars depict the mean and standard deviations of the parameter values of LYSAs inferred by \cite{2025A&A...699A.122O} while the orange lines show samples of the posterior distribution of the linear fit (degree-one polynomial) inferred in this work.}
\label{figure:summary}
\end{figure}

In this work, we identify the most robust statistical correlations in the internal kinematics of LYSAs. Using a recently published catalogue with the linear velocity field parameters and expansion ages of 18 LYSAs, we searched for the most robust (2$\sigma$ credible) correlations using frequentist and Bayesian criteria. Only four correlations among 45 survived both our statistical criteria; these are $\kappa_x$ vs $\kappa_z$, $\kappa_z$ vs $\omega_x$, $\kappa_z$ vs $\gamma_x$, and $\kappa_z$ vs age, which are shown in Fig. \ref{figure:summary}. We estimate a false discovery rate of 53\%, which indicates that at least two of the reported robust correlations could be happening by chance. The confirmation of the true robust correlations will require future confirmatory studies based on independent datasets. The found robust correlations are all related to expansion in the Z direction (i.e. $\kappa_z$ vs age, expansion, rotation, and shear) and, thus, the most probable physical explanation is that they originate due to the gravitational potential of the Galactic disk.

Given the lack of theoretical predictions about the internal motions of sparse stellar systems such as stellar associations, we proposed several explanations for the observed robust correlations. We find that the most probable driving element corresponds to the force exerted by the gravitational potential of the Galactic disk. While this later imprints rotation and shear motions in the parent molecular clouds, expansion results as a by-product of the star-formation process due to the gravitational collapse of these parent clouds. Once the stars are formed, the same Galactic potential that originates the internal motions damps them in shorter timescales than the LYSAs' ages.

The confirmation of this scenario still awaits the arrival of larger samples of stellar systems, particularly those in nearby star-forming regions and large local complexes (e.g. Sco-Cen, Taurus-Auriga). Analysing larger samples of stellar associations, particularly those within the large complexes, will be of paramount importance to disentangle global correlations from those arising at smaller scales due to peculiar star-formation histories. In addition, future work will also be needed to develop more sophisticated models for the observed robust correlations.

\begin{acknowledgements}
We thank the anonymous referee for the useful comments that helped to improve the quality of this work.
JO acknowledge financial support from ‘Ayudas para contratos postdoctorales de investigación UNED 2021’. ‘La publicación es parte del proyecto PID2022-142707NA-I00, financiado por MCIN/AEI/10.13039/501100011033/FEDER, UE’. NMR acknowledges support from the Beatriu de Pinós postdoctoral program under the Ministry of Research and Universities of the Government of Catalonia (Grant Reference No. 2023 BP 00215). P.A.B.G. acknowledges financial support from the São Paulo Research Foundation (FAPESP, grant: 2020/12518-8) and Conselho Nacional de Desenvolvimento Científico e Tecnológico (CNPq, grant: 303659/2024-6).
\end{acknowledgements}

%
\bibliographystyle{aa} 
\bibliography{bibliography} 

\begin{appendix}
\onecolumn
\section{Additional tables}
\label{appendix:tables}
\begin{table*}[h!]
\centering
\caption{\label{table:parameters}Parameters of the polynomial fits of degree zero (P0) and one (P1).}
\resizebox{0.87\textwidth}{!}
{\begin{tabular}{lllrrrrrr}
\toprule
 &  &  & \multicolumn{3}{c}{Intercept} & \multicolumn{3}{c}{Slope} \\
 &  &  & mu & sd & HDI & mu & sd & HDI \\
Correlation & Model & Case &  &  &  &  &  &  \\
\midrule
\multirow[t]{4}{*}{$\kappa_x$ vs $\kappa_z$} & \multirow[t]{2}{*}{P0} & Bayesian & 41.14 & 0.77 & [39.631,42.705] & - & - &   \\
 &  & Frequentist & 41.15 & 0.77 &   & - & - &   \\
 & \multirow[t]{2}{*}{P1} & Bayesian & 50.45 & 1.48 & [47.436,53.371] & 0.21 & 0.03 & [0.155,0.270] \\
 &  & Frequentist & 50.50 & 1.49 &   & 0.21 & 0.03 &   \\
\multirow[t]{4}{*}{$\kappa_x$ vs $\omega_x$} & \multirow[t]{2}{*}{P0} & Bayesian & 41.13 & 0.75 & [39.587,42.608] & - & - &   \\
 &  & Frequentist & 41.15 & 0.77 &   & - & - &   \\
 & \multirow[t]{2}{*}{P1} & Bayesian & 40.76 & 0.96 & [38.909,42.753] & -0.08 & 0.12 & [-0.323,0.175] \\
 &  & Frequentist & 40.78 & 0.95 &   & -0.08 & 0.12 &   \\
\multirow[t]{4}{*}{$\kappa_z$ vs $\omega_x$} & \multirow[t]{2}{*}{P0} & Bayesian & -14.38 & 1.40 & [-17.126,-11.542] & - & - &   \\
 &  & Frequentist & -14.39 & 1.39 &   & - & - &   \\
 & \multirow[t]{2}{*}{P1} & Bayesian & -10.67 & 1.56 & [-13.744,-7.501] & 1.11 & 0.20 & [0.724,1.514] \\
 &  & Frequentist & -10.69 & 1.54 &   & 1.11 & 0.20 &   \\
\multirow[t]{4}{*}{$\kappa_z$ vs $\gamma_x$} & \multirow[t]{2}{*}{P0} & Bayesian & -14.38 & 1.40 & [-17.080,-11.439] & - & - &   \\
 &  & Frequentist & -14.39 & 1.39 &   & - & - &   \\
 & \multirow[t]{2}{*}{P1} & Bayesian & -5.52 & 2.38 & [-10.112,-0.506] & -0.46 & 0.10 & [-0.660,-0.255] \\
 &  & Frequentist & -5.59 & 2.35 &   & -0.46 & 0.10 &   \\
\multirow[t]{4}{*}{$\omega_x$ vs $\gamma_x$} & \multirow[t]{2}{*}{P0} & Bayesian & -4.86 & 1.33 & [-7.483,-2.197] & - & - &   \\
 &  & Frequentist & -4.86 & 1.32 &   & - & - &   \\
 & \multirow[t]{2}{*}{P1} & Bayesian & -3.99 & 1.43 & [-6.870,-1.188] & -0.23 & 0.14 & [-0.529,0.046] \\
 &  & Frequentist & -3.99 & 1.43 &   & -0.23 & 0.15 &   \\
\multirow[t]{4}{*}{$\omega_y$ vs $\gamma_y$} & \multirow[t]{2}{*}{P0} & Bayesian & 0.96 & 0.78 & [-0.569,2.522] & - & - &   \\
 &  & Frequentist & 0.97 & 0.76 &   & - & - &   \\
 & \multirow[t]{2}{*}{P1} & Bayesian & 0.74 & 0.90 & [-1.005,2.526] & 0.07 & 0.14 & [-0.203,0.365] \\
 &  & Frequentist & 0.74 & 0.90 &   & 0.07 & 0.14 &   \\
\multirow[t]{4}{*}{$\omega_x$ vs $\gamma_z$} & \multirow[t]{2}{*}{P0} & Bayesian & -4.87 & 1.31 & [-7.558,-2.263] & - & - &   \\
 &  & Frequentist & -4.86 & 1.32 &   & - & - &   \\
 & \multirow[t]{2}{*}{P1} & Bayesian & -4.15 & 1.40 & [-6.806,-1.294] & 0.25 & 0.14 & [-0.048,0.514] \\
 &  & Frequentist & -4.16 & 1.38 &   & 0.25 & 0.14 &   \\
\multirow[t]{4}{*}{$\gamma_x$ vs $\gamma_z$} & \multirow[t]{2}{*}{P0} & Bayesian & 4.35 & 1.39 & [1.707,7.223] & - & - &   \\
 &  & Frequentist & 4.39 & 1.42 &   & - & - &   \\
 & \multirow[t]{2}{*}{P1} & Bayesian & 4.32 & 1.50 & [1.223,7.244] & -0.02 & 0.14 & [-0.299,0.274] \\
 &  & Frequentist & 4.31 & 1.49 &   & -0.03 & 0.14 &   \\
\multirow[t]{4}{*}{$\kappa_x$ vs Age} & \multirow[t]{2}{*}{P0} & Bayesian & 41.14 & 0.76 & [39.649,42.690] & - & - &   \\
 &  & Frequentist & 41.15 & 0.77 &   & - & - &   \\
 & \multirow[t]{2}{*}{P1} & Bayesian & 49.19 & 4.22 & [40.547,57.594] & -0.25 & 0.13 & [-0.511,0.005] \\
 &  & Frequentist & 49.49 & 4.22 &   & -0.26 & 0.13 &   \\
\multirow[t]{4}{*}{$\kappa_z$ vs Age} & \multirow[t]{2}{*}{P0} & Bayesian & -14.40 & 1.40 & [-17.337,-11.683] & - & - &   \\
 &  & Frequentist & -14.39 & 1.39 &   & - & - &   \\
 & \multirow[t]{2}{*}{P1} & Bayesian & 65.36 & 7.52 & [49.761,79.702] & -2.80 & 0.26 & [-3.312,-2.277] \\
 &  & Frequentist & 66.97 & 7.69 &   & -2.86 & 0.27 &   \\
\multirow[t]{4}{*}{$\omega_x$ vs Age} & \multirow[t]{2}{*}{P0} & Bayesian & -4.88 & 1.31 & [-7.429,-2.216] & - & - &   \\
 &  & Frequentist & -4.86 & 1.32 &   & - & - &   \\
 & \multirow[t]{2}{*}{P1} & Bayesian & -3.21 & 6.92 & [-16.956,10.051] & -0.06 & 0.23 & [-0.522,0.389] \\
 &  & Frequentist & -3.62 & 7.07 &   & -0.04 & 0.24 &   \\
\multirow[t]{4}{*}{$\gamma_x$ vs Age} & \multirow[t]{2}{*}{P0} & Bayesian & 4.41 & 1.43 & [1.646,7.347] & - & - &   \\
 &  & Frequentist & 4.39 & 1.42 &   & - & - &   \\
 & \multirow[t]{2}{*}{P1} & Bayesian & 6.43 & 7.49 & [-9.089,21.067] & -0.07 & 0.25 & [-0.585,0.434] \\
 &  & Frequentist & 6.59 & 7.66 &   & -0.08 & 0.26 &   \\
\multirow[t]{4}{*}{$\gamma_z$ vs Age} & \multirow[t]{2}{*}{P0} & Bayesian & 5.11 & 0.59 & [3.952,6.300] & - & - &   \\
 &  & Frequentist & 5.11 & 0.60 &   & - & - &   \\
 & \multirow[t]{2}{*}{P1} & Bayesian & -0.64 & 6.21 & [-12.246,11.816] & 0.18 & 0.19 & [-0.212,0.539] \\
 &  & Frequentist & -1.08 & 6.28 &   & 0.19 & 0.19 &   \\
\bottomrule
\end{tabular}
}
\end{table*}
\end{appendix}

\end{document}